# Characterizing driver heterogeneity within stochastic traffic simulation

Michail A. Makridis [a*], Aikaterini Anesiadou[b], Konstantinos Mattas[c], Georgios Fontaras[c], Biagio Ciuffo[c*].

[a] ETH Zürich, Institute for Transport Planning and Systems (IVT). Zürich, CH
[b] University of Padova, Department t of Geosciences, Padova, Italy
[c] European Commission – Joint Research Centre, Ispra, Italy

Drivers' heterogeneity and the broad range of vehicle characteristics on public roads are primarily responsible for the stochasticity observed in road traffic dynamics. Understanding the behavioural differences in drivers (human or automated systems) and reproducing observed behaviours in microsimulation attracts significant attention lately. Calibration of car-following model parameters is the prevalent way to simulate different driving behaviors through randomly injected variation around average parameter values. An issue is that, as shown in the literature, most car-following model do not realistically reproduce free-flow acceleration, that is in turn highly correlated with heterogeneity in driving styles. Furthermore, often, model parameters lose their physical interpretation upon calibration. The present study proposes a novel framework to analyse observed vehicle trajectories from various drivers, identify individual driver fingerprints based on their acceleration behavior, cluster drivers in categorically-meaningful driving styles (e.g. mild, normal and aggressive) and reproduce observed individual driver acceleration behaviors in microsimulation. The paper discusses the inability of observed acceleration as a sole quantity to characterise the aggressiveness of a driver and proposes a novel metric called Independent Driving Style (IDS) to perform this task. A large experimental campaign with 20 drivers using interchangeably the same vehicle for one year validates of this work. Finally, simulation results demonstrate the robustness of the proposed method.

# 1 Introduction

Variability in the behaviour of human drivers along and vehicle dynamics is responsible for the emergence of stochastic patterns in the way vehicles move and interact. Traffic flow is composed by many heterogeneous vehicles in terms of power dynamics, powertrains and vehicle manufacturers. Furthermore, stochasticity in vehicle movement under similar conditions is present even for the same vehicle brand and model due to heterogeneous human driving behaviours. On a macroscopic scale, this heterogeneity has been linked to the evolution of several traffic-related phenomena such as capacity drop, traffic hysteresis, traffic oscillations, and stop and go waves, to name a few (Chen et al., 2014; Huang et al., 2018; Laval and Leclercq, 2010; Saifuzzaman et al., 2017; Zheng et al., 2011). Consequently, modeling and reproducing individual driving behaviors within microsimulation attracts a lot of interest in the literature. However, it still remains a very challenging problem.

Several attempts to characterize driver heterogeneity have been carried out in the past decades. In general, research efforts have so far focused on the variability in the behavior of the same driver, called intra-driver heterogeneity (Berthaume et al., 2018; Huang et al., 2018; Wang et al., 2010), or of different drivers, called inter-driver heterogeneity (Taylor et al., 2015), or both (Ossen and Hoogendoorn, 2011). Heterogeneity has been linked to several factors such as the driver personal information (gender, age, etc.) (Yagil, 1998), weather conditions (Kilpeläinen and Summala, 2007), traffic conditions (Lajunen et al., 1999), mood (Underwood et al., 1999) and cross-cultural differences (Özkan et al., 2006). Several attempts have been made to cluster similar driving traits. This has usually involved naturalistic driving data collection from experiments and/or information extraction using statistical learning methods (Al Haddad and Antoniou, 2022).

With the advent of automated driver assistance systems, such as Adaptive Cruise Control (ACC), a reduction (at least) in drivers' heterogeneity and most probably also in the variety of the observed vehicle dynamics is expected (Makridis et al., 2021). This gives an additional motivation to gain a deeper understanding and accurate description of observed driving behaviour, as this would facilitate meaningful comparative studies between human and automated drivers (Calvert and van Arem, 2020; Zhao et al., 2020).

Within microscopic traffic models, car-following models reproduce the longitudinal driving task. Different driver behaviors are captured stochastically, injecting white noise on the model's acceleration parameter (Laval et al., 2014; Ngoduy et al., 2019; Treiber et al., 2006; Treiber and Kesting, 2013). However, such implementations omit any physical interpretation for the calibrated model parameters, e.g. the variation in the acceleration model parameter between drivers, most probably, does not reflect the variation in the observed acceleration between those drivers in the real world. Additionally, in most car-following models, realistic simulation of free-flow acceleration dynamics is neglected (Ciuffo et al., 2018), despite the fact that an increasing number of works has recently highlighted the importance of free-flow acceleration even under congested traffic conditions (Laval et al., 2014), with the emergence of free-flow pockets (Makridis et al., 2020). Building on the findings of these recent studies, the present paper also proposes a novel way to simulate driver heterogeneity by stochastically integrating it in the MFC model (Makridis et al., 2019), which explicitly accounts for the dynamics of vehicle powertrain. A Python implementation of the MFC model for the simulation of internal combustion engine vehicles is openly available online[1.] This model has been recently incorporated in Aimsun Next[2] and therefore the methodology presented in the paper can be easily used in traffic simulation studies by practitioners.

---

[1] https://pypi.org/project/co2mpas-driver/
[2] https://www.aimsun.com/aimsun-next-new-features/

Usually, driving behaviour can be influenced by several stochastic factors (intra-driver heterogeneity), such as drivers' mood, the time of the day, the traffic state, the weather and many others (Hoogendoorn et al., 2011). So, the ability to create a model that captures a driver's behaviour, assuming to be able to properly characterize it, can be achieved only through a stochastic process characterized by an average behaviour and an expected variance. In microsimulation, as mentioned also above, acceleration is used as a receptor parameter for the above-described stochasticity. However, we argue that acceleration is not an objective metric to characterise a driver. The reason is that the power dynamics are not uniform across the speed range of a vehicle. Calibration of driver behaviour on observed acceleration includes also the employed vehicle dynamics, i.e. it is more calibration of the vehicle-driver system. Hence, considering only the acceleration observations without taking into account the instantaneous speeds can produce misleading conclusions for the driver's aggressiveness. For example, let us imagine an aggressive driver accelerating from an initial speed of 10km/h and the same vehicle-driver system accelerating from an initial speed of 100km/h. In the first case, the observed instantaneous accelerations will probably be much larger than in the second case. The reason is that the vehicle's nominal acceleration capacity reduces with the increase in the speed. In this work, we therefore also discuss about the inability of acceleration as a metric to describe the driver aggressiveness in different speeds and thus to provide fair quantitative comparisons between drivers that drive in different traffic conditions. To solve this issue we introduce a novel metric that considers acceleration as a function of speed to facilitate vehicle-independent and speed-independent driver comparisons.

First, the present work analyses both inter and intra-driver heterogeneity in a concise way, providing a robust indicator for quantitative comparisons between different drivers. Second, we propose a novel way to simulate driver heterogeneity in microsimulation environments. Third, introduce a novel driver characterization metric that considers acceleration as a function of speed to facilitate vehicle-independent and speed-independent driver comparisons.

The paper is structured as follows: Section 2 presents the proposed framework, the updated model and the experimental campaign, Section 3 presents the results and Section 4 the conclusions.

## 2 Methodology

We propose a methodological approach to analyse a driver's behaviour based on observed vehicle trajectories and reproduce observed behaviors in microsimulation. A high-level schematic representation of the proposed framework is illustrated in Figure 1. The upper part refers to the driver behaviour analysis based on observations. The lower part reproduces characterized behaviors in microsimulation. The different components are discussed in the next sections.

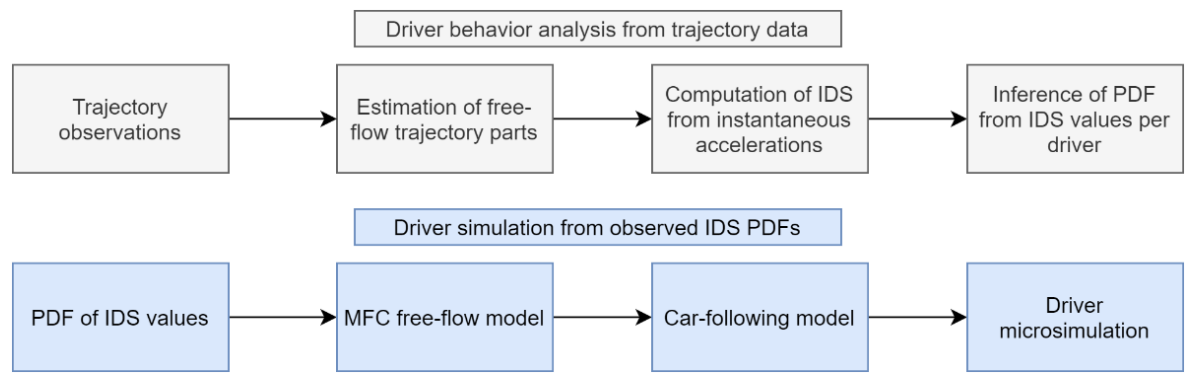

Figure 1. High-level schematic representation of the proposed framework. The top part refers to driver behaviour analysis; the lower part to driver behaviour simulation.

### 2.1 Independent Driving Style (IDS)

The behavioural differences between drivers are more prominent when other drivers or obstacles do not constrain them, that is under free-flow conditions or when leading vehicles are far away. This information is usually not recorded during experiments and thus not available for analysis.

The following section describes how we estimated it.

*2.1.1 Estimation unconstrained vehicle movements*

We adopt a simple indirect way to estimate unconstrained vehicle movement, through short trajectory parts with sharp accelerations. The idea is that under congested conditions the vehicles are not able to accelerate sharply, as they are bounded by the leading vehicles. From trajectory data, it is relatively easy to detect acceleration events in which a vehicle accelerates from a starting speed to a higher one (with a speed jump of $\Delta v$). If the duration of this acceleration event, namely $w$, is short enough, while and the speed jump $\Delta v$ is large enough (indicating sharp acceleration), then we can do a logical assumption that this vehicle is moving unconstrained. The higher the $\Delta v$, the longer the acceptable $w$ can be.

First we apply a simple local min/max algorithm to detect all possible acceleration events (matching sequential local minima and local maxima). Then, we need to identify those who correspond to unconstrained movement based on proper thresholding of parameters $\Delta v$ and $w$. Fortunately, we can fine-tune these two parameters through publicly-available car-following data under the "reductio ad absurdum" logic (Makridis et al., 2021). In those experiments, vehicles follow their leader in platoon formation. During the driving cycle, the leader randomly acceleration and decelerates sharply and occasionally the movement of the following vehicles can be considered unconstrained.

In the above dataset, we detect all acceleration events recording parameters $\Delta v$ and $w$. Based on short, medium, long and very long duration we set a threshold value for $\Delta v$ the 75$^{th}$ percentile of recorded values and we distinguish constrained versus unconstrained movements. An example of estimated free-flow acceleration events is shown in Figure 2.

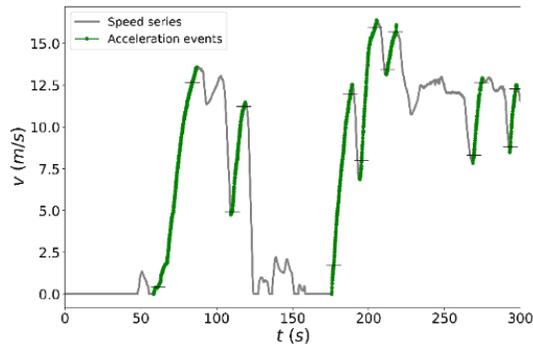

Figure 2. Identified sharp acceleration events corresponding to unconstrained vehicle movement in experimental observations.

*2.1.2 Acceleration as a measure for driver characterisation*

The instantaneous acceleration is often used in the literature to characterise drivers' style (Berthaume et al., 2018; Hamdar et al., 2015; Hoogendoorn et al., 2011; Taylor et al., 2015). However, instantaneous acceleration can be misleading for comparative analyses when the drivers use vehicles with very different power capabilities or the observations refer to different traffic conditions or driving environments. To explain this concept, Figure 3 has been derived through modelling from available experimental data and describes the acceleration over speed domain for one specific vehicle. In particular, in Figure 3 the cyan line depicts the acceleration potential of the vehicle across the whole speed range. The yellow line shows the maximum common acceleration values observed at each given speed, and similar the orange line shows the minimum common deceleration values observed at each given speed for the specific vehicle. These three lines define four regions, namely Regions A, B, C and D.

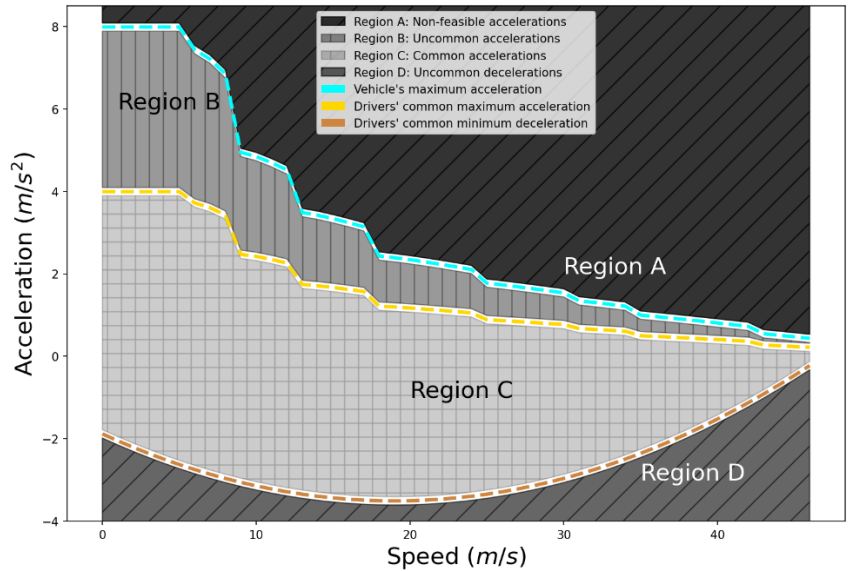

Figure 3. The unrealistic, feasible and common acceleration-speed domain (model representation) for the vehicle used in the experimental campaign described in Section 2.4.

Region C contains most frequently observed acceleration values. The union between Region B and C describes the domain of plausible values, while Regions A and D include unrealistic acceleration values that the vehicle cannot deliver due to power or other limitations. From the figure, it is possible to understand the unsuitability of acceleration to characterise driving aggressiveness for the following reasons. First, when two vehicles have a significant difference in power capabilities, their acceleration at a given speed is not comparable. Second, a driver adapts to the vehicle capabilities. Therefore, it is unfair to characterise a driver of a sport car as more aggressively only because the car can produce higher acceleration than conventional ones. Third, the acceleration response of commercial vehicles is not uniform across their speed range. It is obvious from Figure 3 that the maximum possible acceleration for speeds around $10 m/s$ (urban) is higher than $4 m/s^2$, while for speeds around $30 m/s$ (freeway) is around $1 m/s^2$. It is therefore unfair to compare drivers observed with acceleration values measured on high speeds with those measured on low speeds.

To address this problem, in the present work we introduce using vehicle dynamics modelling tools, an acceleration-based metric that is independent of the vehicle power and the current speed. We achieve this in a two-step methodology described in the next sections.

*2.1.3 Vehicle-independent acceleration values*

Transformation of instantaneous acceleration values to vehicle-independent ones is possible if we know the vehicle specifications. Assuming known vehicle specifications it is indeed possible to define the maximum possible acceleration that that vehicle can deliver at a given speed and by using this value the acceleration can be described as a rate of the vehicle's acceleration potential at the given speed.

There are several models of this type in the literature (Fadhloun and Rakha, 2020; Makridis et al., 2019; Rakha et al., 2004, 2012). Here, we use the MFC model proposed in (Makridis et al., 2019). The employed model has been extended also for electric vehicles that demonstrate different power distribution across speed range (He et al., 2020).

Using the MFC model we construct the acceleration capacity function of the vehicle, namely $a_{cp}(v)$ (i.e. cyan line in Figure 3), where $v$ is the instantaneous speed. It should be noted that $a_{cp}(v)$ is also a function of the current gear, which is computed based on the work of (Fontaras et al., 2018). Then we express the instantaneous acceleration $a$ as a rate of $a_{cp}(v)$:

$$ds = \frac{a}{a_{cp}(v)} \quad (1)$$

As it can be perceived, the $ds$ quantity is related to the acceleration but it is independent of the vehicle power capabilities. In generalised studies, the reader can use average acceleration potential functions based on average vehicle performance in commercial fleets to avoid the need for knowledge of each individual vehicle's specifications.

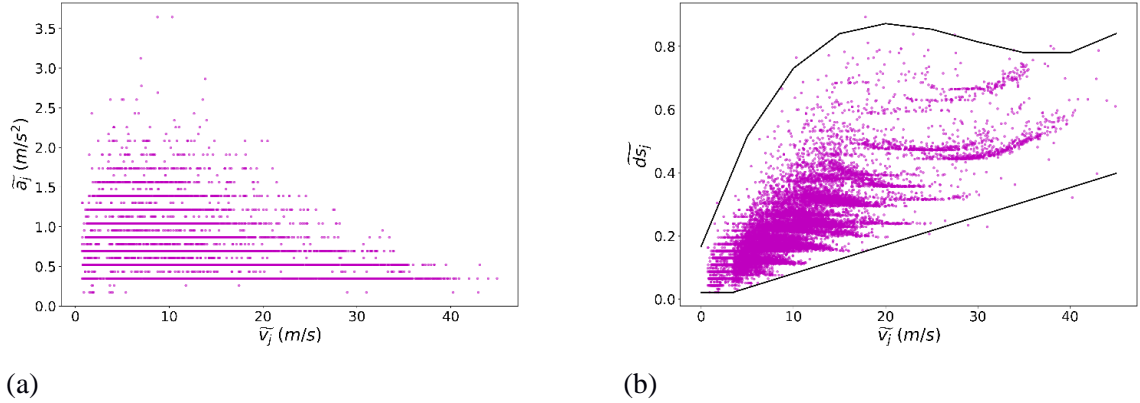

(a)                                        (b)

Figure 4. Naturalistic data refer to the 20 observed drivers of the experimental campaign used in this work. Fig. 4 a) shows acceleration over speed per movement, and b) the proposed $ds$ values per speed per movement.

In the remaining part of the driver analysis, we perform the logical assumption that a driver does not change his/her driving behaviour during a single acceleration event. Therefore, instead of working on instantaneous acceleration, we continue our analysis based on the median acceleration/speed values observed during a detected unconstrained movement. This is a logical assumption that reduced computational complexity and provides invariability to noisy observations (Punzo et al., 2011).

Figure 4 is an intuitive demonstration of this new vehicle-independent metric. The plots refer to naturalistic data from a large pool of drivers (see later the description of the experimental campaign). Each point in Figure 4a shows the median acceleration and speed values of an observed unconstrained movement. As expected, the observed acceleration values are much higher in low speeds than in high speeds due to non-uniform vehicle power dynamics. Figure 4b shows the proposed $ds$ values and speed values. The black lines are piece-wise linear functions used to inscribe the operational domain of all observed values.

*2.1.4 Speed-independent acceleration values*

We discussed how $ds$ is a vehicle-independent metric but the correlation with speed remains. The observed $ds$ values are lower in low speeds and higher in high speeds, demonstrating opposite tendency than acceleration values. We employ a simple normalization process to tackle this issue. First we derive two piece-wise equations that describe the black lines observed in Figure 4b, namely $f_{min}(v)$ and $f_{max}(v)$. The fitted curves are reported in Appendix Part A for reproducibility of the proposed study. Second, we express the observed $ds$ values as a rate of the operational domain that $f_{min}$ and $f_{max}$ describe as follows:

$$ids = \frac{ds(v) - f_{min}(v)}{f_{max}(v) - f_{min}(v)} \tag{2}$$

Where $ids$ is called Independent Driving Style (IDS), $ds$ comes from Eq.1 and $v$ is the observed median speed in an unconstrained movement. For the sake of demonstration in this work, we assume that the $f_{min}(v)$ and $f_{max}(v)$ have global validity. In other words, we assume that our pool of drivers is large enough to describe all possible driving behaviors, or our dataset is large-enough to capture all possible networks and traffic conditions. Consequently, in real-world application, the $f_{min}(v)$ and $f_{max}(v)$ need to be updated as new observations arrive in order to describe the drivers' operational domain properly. Figure 5 shows that normalized $ids$ values per speed.

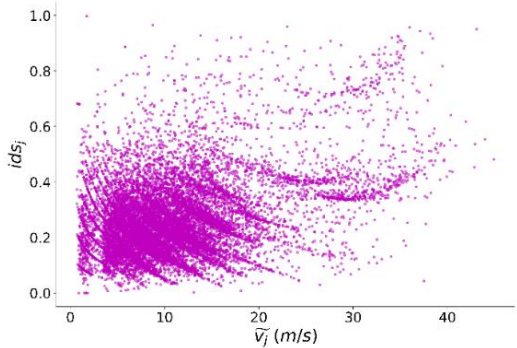

Figure 5. Uniform distribution of IDS values across the whole speed range. Data refer to 20 observed drivers ($j$ denotes the ID of an acceleration event).

*2.2 Driver characterisation using distributions of IDS values.*

Each observed IDS value corresponds to an observed unconstrained vehicle movement and the main assumption is that this movement can be part of the driver's fingerprint and characterize him. Assuming enough observations per driver, we can construct the IDS histogram of this driver and then approximate the derived histogram with the probability density function (PDF) to enable analytical comparisons. In this work, we assume and validate (later in the paper) our assumption that driver IDS observations follow a lognormal PDF. Each fitted PDF can be thought as the driver's behavioural fingerprint.

*2.3 Stochastic microsimulation of observed drivers*

Assuming a composed IDS PDF for a driver this section describes how proposed simulation of this driver can be achieved. The IDS PDF is used to randomly select the acceleration of a vehicle in a Monte Carlo fashion. In particular, a randomly sampled number in the [0,1] is thus first defined. Via the PDF the number is translated into a value of IDS and using the following two simple equations (which reverse Eq. 1 and 2) into an acceleration value:

$$ds = ids \cdot (f_{max}(v) - f_{min}(v)) + f_{min}(v) \quad (3)$$
$$a = [ds \cdot (f_{max}(v) - f_{min}(v)) + f_{min}(v_t)] \, a_{cp}(v) \quad (4)$$

It is worth underlying that once the efforts to obtain the IDS PDFs have been made, using the model in combination with traffic simulation becomes extremely simple and light.

*2.4 Experimental campaign*

In order to validate the proposed methodology we conducted an experimental campaign in which 20 different drivers used the same vehicle under different driving conditions for a period of 12 months. The vehicle used in the test campaign is a 2.0L diesel C-segment passenger car, equipped with automatic transmission (9 gears), leased for 12 months (during 2016 and 2017). The dataset consists of more than 10.000 km driven by 20 drivers using the same vehicle. During the experimental campaign the vehicle was assigned to a volunteer driver, who was free to use it as a personal vehicle for a period of approximately 2 weeks without any restriction on the frequency of usage, the driving style, the fuel consumption or the route choice. For additional information on the test campaign, the reader can also refer to the corresponding publication (Pavlovic et al., 2020).

Table 1 provides an overview of the dataset with distance and trips' time duration per driver. Time duration refers to actual driving time without long pauses. The differences in the number of driven kilometres or hours travelled can be explained by the fact that there was no restriction or guidance on how each driver uses the vehicle. Additionally, this table shows the number of acceleration events detected for each driver, along with the distance in km covered during those events.

It is worth noting that the distance travelled by the 20 drivers is not the same. Therefore, the statistics presented in Table 1 vary among the drivers. Since a main assumption in the paper is that the collected data per driver are enough to accurately describe his driving profile, this difference may affect the generality of the results achieved. Nevertheless, we do not consider this limitation critical for demonstrating the robustness of the proposed approach and additionally we expect that it will disappear in the future with the abundance of in-vehicle (e.g. telematics) or infrastructure (e.g. Bluetooth) sensors that can collect large volumes of personalized driver data.

Table 1. Basic statistics of the trips per driver.

| Driver | Total kms | Total hours | Free-flow events number | Events (km) | Driver | Total kms | Total hours | Free-flow events number | Events (km) |
|--------|-----------|-------------|-------------------------|-------------|--------|-----------|-------------|-------------------------|-------------|
| D1 | 250 | 8 | 715 | 117 | D11 | 1056 | 14 | 420 | 171 |
| D2 | 669 | 9 | 447 | 129 | D12 | 119 | 4 | 456 | 55 |
| D3 | 1973 | 39 | 2807 | 538 | D13 | 1039 | 21 | 1401 | 307 |
| D4 | 300 | 4 | 223 | 46 | D14 | 1189 | 20 | 1009 | 295 |
| D5 | 1055 | 14 | 482 | 134 | D15 | 311 | 8 | 548 | 122 |
| D6 | 45 | 2 | 176 | 20 | D16 | 2581 | 35 | 942 | 405 |
| D7 | 166 | 3 | 151 | 58 | D17 | 288 | 7 | 632 | 118 |
| D8 | 442 | 13 | 1054 | 150 | D18 | 183 | 7 | 614 | 77 |
| D9 | 310 | 7 | 619 | 111 | D19 | 303 | 6 | 345 | 68 |

| D10 | 174 | 6 | 481 | 71 | D20 | 1424 | 19 | 1139 | 297 |

## 3 Results

This section presents the results of the proposed methodology to characterize intra- and inter-driver heterogeneity, as well as simulate specific driver profiles.

### *3.1 Inferring the PDF distribution per driver*

In the methodology section, IDS was identified as a normalised metric, independent of the speed and the vehicle's powertrain, necessary to perform an objective comparison between different drivers. The IDS values per driver were derived from the median accelerations observed during the acceleration events derived using the methodology described in Section 2. The derivation of lognormal PDF was assessed using the Kolmogorov-Smirnov (K-S) test, which is a nonparametric statistical test of the equality of one-dimensional probability distributions. K-S is used here as a goodness of fit test to compare the observed histogram with the continuous reference-fitted PDF.

In Figure 6, the empirical distributions of observed IDS values per driver are presented with the y-axis denoting probability density. The lognormal distribution has been indeed mentioned in the literature as suitable for the description of human behaviour. Gualandi and Toscani studied the connection between human behaviour and lognormal distribution also considering cases of modelling drivers in traffic (Gualandi and Toscani, 2019). Furthermore, Antoniou et al. (Antoniou et al., 2002) concluded that the aggregation of traffic measurements forms a statistical distribution, which is quite accurately described by a lognormal distribution. The goodness of fit test results based on the K-S method are reported in the Appendix Part B. The parameters of the lognormal PDFs for all drivers are given for reference in Appendix Part C.

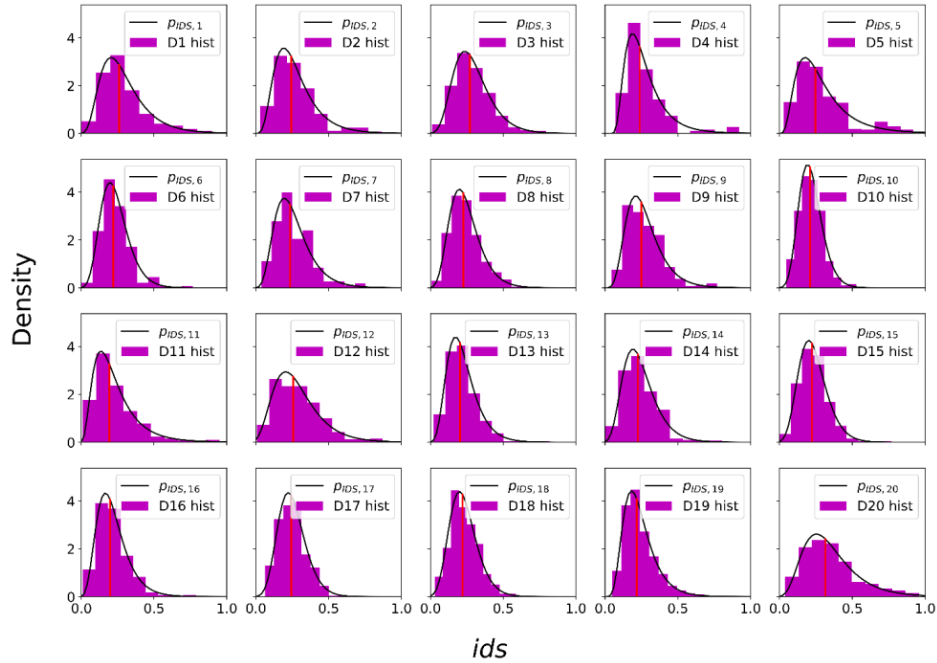

Figure 6. Histogram of the observed IDS values for each driver along with the fitted lognormal PDF. The vertical red lines denote the location of the median.

## 3.2 Inter-driver heterogeneity

With inter-driver heterogeneity we refer to the difference in the driving styles among drivers. This heterogeneity is explored through the fitted PDF related to each driver. A visual inspection of the different driving behaviours is already visible from Figure 6. To have a more marcoscopic representation of the inter-driver heterogeneity, we perform a categorical clustering for the 20 drivers into conceptually-perceived driver profiles, i.e. mild, normal or dynamic. We attempt to cluster all drivers in the afore-mentioned generic categories using the simple K-means algorithm (Macqueen, 1967). The variables used for the K-means clustering refer to descriptive statistics of the lognormal PDFs. Three features used for that, one related to the central tendency of the PDF and another two related to the spread of the PDF (25th, 75th percentiles).

Figure 7 (a) presents the results of this grouping in the form of a 3D plot. The axes refer to the median, the 25th and the 75th percentile values. Drivers belonging to the same group are plotted with the same marker, while black points show the clustering centers, which are the representatives of each group. The results are clearly bounded by the small size of the dataset, yet interesting findings can be deduced. One driver (D20) exhibited significant differences to the rest of the drivers, which was also obvious in the visual inspection and constituted a separate group. Figure 7 (b) presents the distribution per driver category. As expected, there is a large overlapping area between the different driver clusters. An aggressive driver is not expected to drive constantly in an aggressive way, as this is not always allowed, e.g. due to traffic state, weather, mood or other stochastic factors. The distribution of mild drivers shows higher probability for lower IDS values than normal or aggressive ones. For reference, later in the paper, Drivers with ids 6,7,10,11,13,15,16,18 and 19 are clustered as mild, drivers with ids 1,2,3,4,5,7,9,12 and 17 are clustered as normal and driver with id 20 is alone in aggressive cluster.

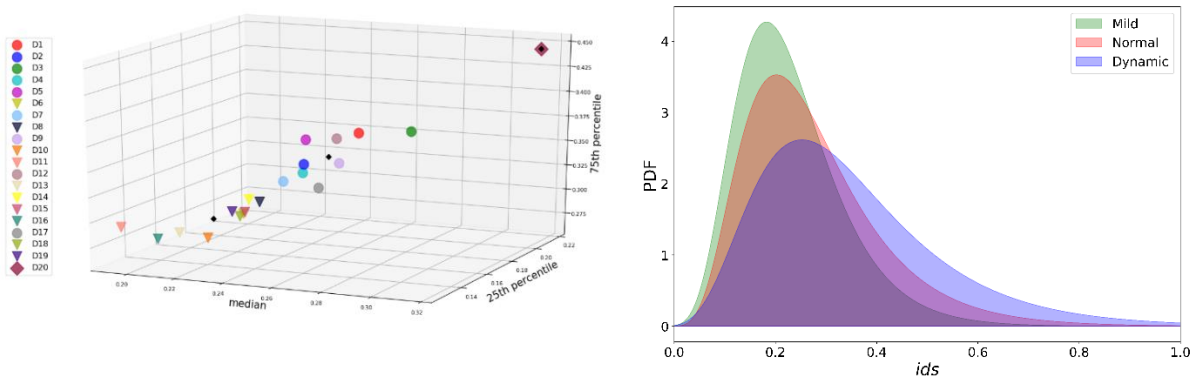

Figure 7. a) K-means clustering result presented in a 3D plot, b) PDFs of the inferred mild, normal and dynamic driver type.

*3.3 Intra-driver heterogeneity*

Intra-driver heterogeneity refers to the different driving styles that a driver employs under various conditions. This kind of heterogeneity is assessed here since it is a crucial factor potentially affecting the observed variation in traffic-related variables such as headway (Li et al., 2020). An idea of the variation on the driving style of a single driver, is already taken by observing the fitted PDF. Its wideness visually shows which driver is more consistent with his driving style with respect to others.

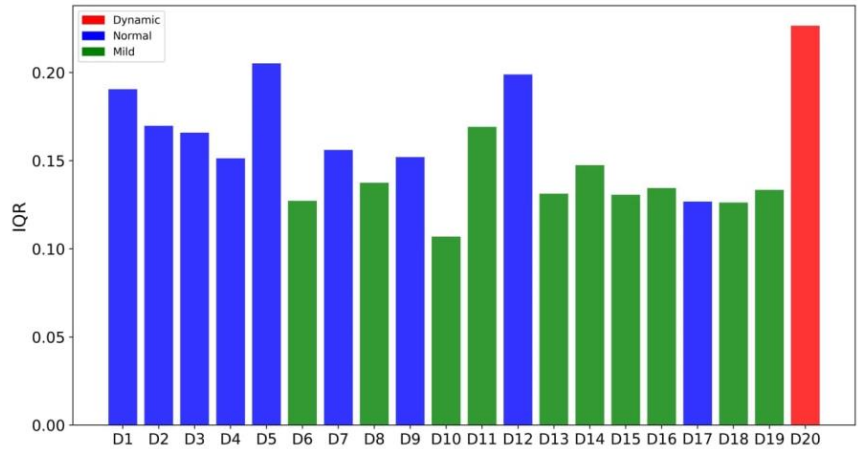

Figure 8. Bar plot of the IQR of the drivers' PDFs.

To quantitatively explore the intra-driver heterogeneity, we refer again to descriptive statistics. More specifically, we use a measure of statistical dispersion, the interquartile range (IQR), which is equal to the difference between 75th and 25th percentiles. Drivers with bigger IQR are considered to have a bigger degree of intra-driver heterogeneity. This heterogeneity is linked only to the different driving behaviours since the vehicle remains the same in all cases. Figure 8 illustrates the IQR values for each driver. Drivers D1,D5,D12 (normal cluster) and D20 (dynamic cluster) demonstrate the biggest IQR, and therefore largest intra-driver heterogeneity. Furthermore, as expected most mild drivers demonstrate low intra-driver heterogeneity, i.e. narrow IDS distributions. These quantitative results are aligned with expected empirical behaviours for mild, normal and dynamic drivers.

### 3.4 Simulation of individual driving styles

The individual driving styles are commonly simulated through stochasticity in the acceleration parameter of the employed car-following model. However, the random variance includes aggregation of vehicle dynamics, driver aggressiveness and possibly other factors as well. Therefore, upon calibration, it most probably loses its physical representation, i.e. maximum acceleration. For simulation of individual driver profiles, we suggest detailed models that introduce explicit description of vehicles dynamics and driver profile (Fadhloun and Rakha, 2020; Makridis et al., 2019). This work uses the Microsimulation Free-flow aCceleration model (MFC) towards this purpose (Makridis et al., 2019). In a previous study by the authors, it was demonstrated how the MFC can be incorporated in car-following models by replacing their free-flow component (Makridis et al., 2020). Results have shown that adopting a simplified Lagrangian Godunov scheme (Leclercq et al., 2007) the MFC is able to reproduce macroscopically-observed phenomena with the formation and propagation of traffic oscillations in the absence of lane changes that can be explained by the stochastic nature of drivers' acceleration processes. Moreover, MFC supports the car-following model to reproduce the concave growth pattern of the speed standard deviation for a group of vehicles in car-platoon formation. Figure 9 illustrates the MFC behavior within the framework of the LWR model.

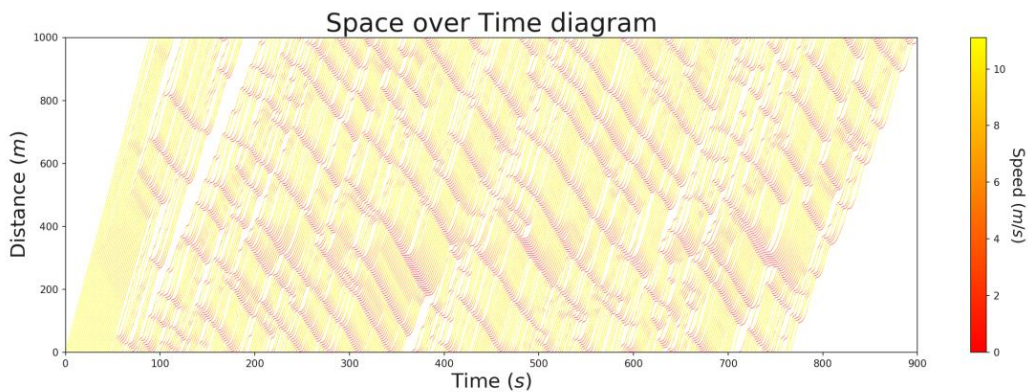

Figure 9. Reproduction of empirically-observed traffic oscillations with the MFC-LWR model as shown in (Makridis et al., 2020).

In the present work, instead of using a constant driver behaviour parameter, we sample on real time the driver behaviour parameter from the driver's derived distribution (PDF). Since the aim of this work is not on traffic oscillations, in this section, we provide validation results only on the new free-flow behaviour of the proposed model. However, it can be expected that the new model can produce oscillations of even more variable amplitudes and frequencies.

The assessment of the proposed model to simulate different drivers is performed in two levels. First, all the events with observed unconstrained movements from all drivers are considered and simulated, i.e. acceleration from the low speed to the high speed of a certain event. Model-wise, each event is simulated based on a randomly sampled IDS value from the respective PDF of each driver. For each event, we perform 10 simulations, each one with a different random IDS sample. Figures 10 shows the observed and the simulated driver distributions for median acceleration values per unconstrained movement, for all drivers. The results are very promising and it is worth noting that close reproduction of speed and acceleration values per driver can help towards more accurate simulation of the driver's energy consumption profile as well.

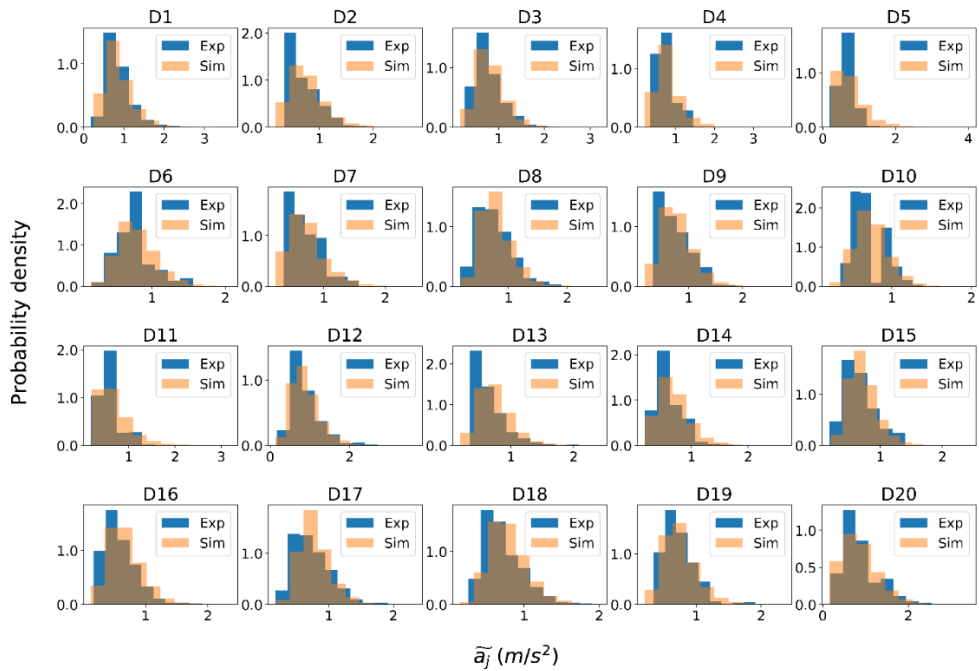

Figure 10. Median acceleration per event histograms of measured and simulated data.

Looking at individual unconstrained movements, for each event of each driver, we perform two simulations, using the 25$^{th}$ and 75$^{th}$ percentiles from the driver's PDF. Figure 11 shows for an indicative event per driver how empirical accelerations lie within the area inscribed by the two simulations.

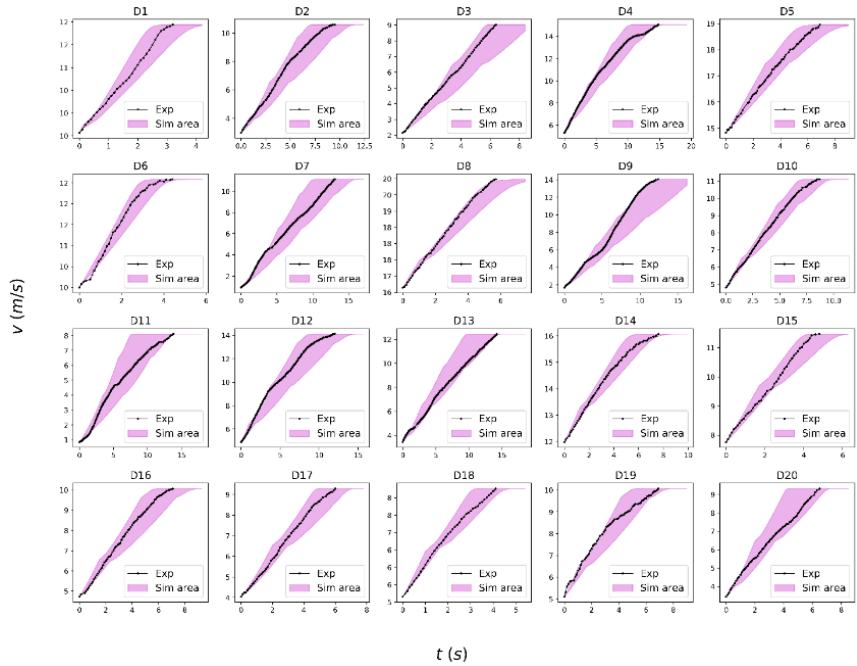

Figure 11. An indicative unconstrained movement per driver. For each event, the simulation area with the 25$^{th}$ and 75$^{th}$ percentile from the driver's PDF is highlighted.

Finally, it is interesting to see how different driver clusters appear in simulation. We setup an acceleration scenario with a fixed duration of 90s from 0 to 30m/s desired speed with the proposed model. If the driver reaches the desired speed before the end of the scenario duration, the same speed is kept for the remaining time. We perform 1000 simulations with randomly selected IDS values from each cluster's PDF, i.e. 3000 simulations in total. Figure 12 shows histograms of median accelerations for the simulated event per driver type. One can distinguish the different types of drivers but also the overlapping areas in the respective PDFs, as expected in real-world observations. Finally, looking from a high-level analysis point of view, we see in the majority of the cases, the dynamic driver is faster than the normal (61.5% - cases 1,2,6), and the mild one

(70.1% - cases 1,2,3) and the mild driver is slower than that dynamic (70.1% - cases 1,2,3) and the normal one (59.5% - cases 1,3,4).

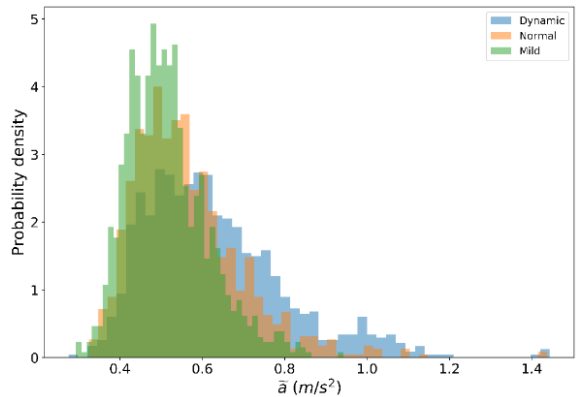

Figure 12. Histograms of median accelerations of the simulations concerning the three driver types.

## 4 Discussion and conclusion

Variability in the behaviour of human drivers along and vehicle dynamics is responsible for the emergence of stochastic patterns in the way vehicles move and interact. On a macroscopic scale, this heterogeneity is responsible for the appearance of various traffic-related phenomena. Thus, accurate modeling and reproduction of individual driving behaviors within microsimulation attracts a lot of interest in the literature. Car-following models reproduce the longitudinal driving task. Different driver behaviors are captured stochastically, injecting white noise on the model's acceleration parameter. While this can be sufficient for simulation under car-following conditions (when there is a leader ahead), this seems not the case for free-flow acceleration conditions. Free-flow parts in most car-following model are simplified and do not reproduce realistic acceleration dynamics. This under-representation of acceleration affects not only free-flow but also saturated traffic conditions when free-flow pockets arise between the vehicles.

The gaps that currently exist in the literature and the proposed paper aims to tackle can be summarized as follows:

- Weak description of free-flow acceleration in microsimulation.

- Unfair comparison of drivers' acceleration behaviour based on instantaneous accelerations.

- Challenging to reproduce an observed driver profile in microsimulation.

- Challenging to aggregate different driver behaviors in driver cluster representations based solely on trajectory data.

- High gaps between observed and simulated driver emission estimates.

The present study proposes a novel framework to characterise driver acceleration behaviour based on the analysis of vehicle trajectories. It discusses why instantaneous acceleration can be misleading for comparative analyses when the drivers use vehicles with very different power capabilities or the observations refer to different traffic conditions or driving environments. It proposes a vehicle- and speed-independent acceleration-based metric, namely the Independent Driving Style (IDS). Afterwards, based on empirical trajectory data from 20 drivers, it studies distributions of IDS values per driver and proposes a concrete way of stochastic driver simulation based on IDS probability density functions.

The contributions of this paper are summarized as follows:

- Clarifying of the role of instantaneous acceleration on driving behaviour representation and comparison among drivers.

- Develop of a vehicle- and speed-independent indicator, (IDS) to describe the driver acceleration behaviour.

- Demonstrate of IDS for the profile description of individual drivers or driver clusters, i.e. mild, normal, dynamic.

- Reproduce accurately observed individual driver profiles in microsimulation.

Although the employed driver dataset is one of the largest publically available ones in such high measurement frequency, it is still rather small to describe all the possible driver profiles under real-world conditions. This poses limitations on the proposed framework that constitute the present results preliminary. More specifically, it is still not clear the amount of individual driver data needed for a complete PDF. Additionally, domain descriptive piece-wise linear functions, $f_{min}$ and $f_{max}$, introduced in Section 2.1.4 will change as more data will be used. Finally, categorical clustering of drivers is used only indicatively to highlight the possibilities of the proposed framework.

Future work will assess through large-scale experimental data, the capability of the proposed study to describe and differentiate the behavioural differences of automated driver assistance systems, as well as compare their behaviour with human drivers.

## Acknowledgements

This research has been funded by the Joint Research Centre of the European Commission. The views expressed are purely those of the authors and may not, under any circumstances, be regarded as an official position of the European Commission. The work by A. Anesiadou has been carried out at the European Commission Joint Research Centre. The authors are grateful to Vincenzo Arcidiacono for the help regarding CO2MPAS gear identification tool and to Jelica Pavlovic and Kostis Anagnostopoulos for the help regarding the pre-processing of the experimental data. The views expressed in the paper are purely those of the authors and may not, under any circumstances, be regarded as an official position of the European Commission

# Appendix

### Part A - Fitted curves defining the operational domain in median *DS*-median Speed space

Table A.1 Characteristics of fitted curves along with the special conditions for each one.

| Curve | Polynomial curves | Special conditions |
|---|---|---|
| $fmin$ | $fmin(\tilde{v}) = 0.009 \cdot \tilde{v} - 0.009$ | $fmin(\tilde{v}) \geq Min\ \widetilde{ds}_{exp}$ |
| $fmax$ | $fmax(\tilde{v}) = 3.70 \cdot 10^{-5}\ \tilde{v}^3 - 0.003\ \tilde{v}^2 + 0.084 \cdot \tilde{v} + 0.167$ | |

Where $Min\ \widetilde{ds}_{exp}$ is the minimum observed median $ds$ value equal to 0.021, $\tilde{v}$ is the median speed value.

### Part B - Goodness of fit K-S test of all drivers

Table B1 shows the results of the goodness of fit K-S test of all the $p_{IDS,i}$ functions. In all driver cases, the test statistic value was smaller than the critical value ($D < Critical$) which allows us not to reject the null hypothesis and accept, with a 1% level of significance, that all $ids_{i,j}$ distributions can be represented by the inferred $p_{IDS,i}$s. Additionally, the resulting p-values were larger than the significance level (p-values ≫ significance level), providing more evidence on not rejecting the null hypothesis.

Table B.1 The goodness of fit test results based on the K-S method.

| Driver | Critical value | D statistic | p-value | Driver | Critical value | D statistic | p-value |
|---|---|---|---|---|---|---|---|
| D1 | 0.061 | 0.037 | 0.280 | D11 | 0.080 | 0.019 | 0.999 |
| D2 | 0.077 | 0.032 | 0.739 | D12 | 0.076 | 0.036 | 0.596 |
| D3 | 0.031 | 0.013 | 0.710 | D13 | 0.043 | 0.019 | 0.662 |
| D4 | 0.109 | 0.042 | 0.822 | D14 | 0.051 | 0.032 | 0.264 |
| D5 | 0.074 | 0.040 | 0.404 | D15 | 0.070 | 0.023 | 0.928 |
| D6 | 0.123 | 0.067 | 0.391 | D16 | 0.053 | 0.035 | 0.188 |
| D7 | 0.132 | 0.048 | 0.870 | D17 | 0.065 | 0.026 | 0.768 |
| D8 | 0.050 | 0.015 | 0.969 | D18 | 0.066 | 0.024 | 0.859 |
| D9 | 0.065 | 0.036 | 0.396 | D19 | 0.088 | 0.035 | 0.782 |
| D10 | 0.074 | 0.039 | 0.435 | D20 | 0.048 | 0.022 | 0.662 |

**Part C - Parameters of the lognormal PDFs of all drivers**

A three-parameter lognormal PDF $p_{IDS,i}$ (Singh, 1998), with $i$ being the driver id, is fitted at each one of them and shown as a black curve in the figure. The first parameter (shape) denotes the general form-shape of the distribution and is also equal to the standard deviation of the $ln(ids_{i,j} - location)$, the second one (location) reveals the shift on the x-axis, representing a lower bound and the last one if combined with the shift parameter (scale + location) expresses the median of the $p_{IDS,i}$. For completion, Appendix B includes the parameters of the inferred $p_{IDS,i}$s. The location of the median for each $p_{IDS,i}$ is denoted with a vertical red line in the figure.

Table C.1 presents the lognormal parameters which actually characterise the drivers. One can see that in most cases location parameters are slightly negative which means there is a tiny probability that the PDF takes negative values (more specifically, among all probabilities for an $p_{IDS,i}$ to get values smaller or equal to zero, the biggest one is 0.17%). At this stage we also note that those parameters do mathematically reconstruct each $p_{IDS,i}$, however, they do not directly reveal meaningful information for the lognormal itself, apart from the location shift. Therefore, descriptive statistics of each $p_{IDS,i}$ are used to analyse further the drivers' characteristics.

Table C.1 Parameters of the lognormal PDFs.

| Driver | shape | location | scale | Driver | shape | location | scale |
|---|---|---|---|---|---|---|---|
| D1  | 0.448 | -0.051 | 0.311 | D11 | 0.560 | -0.025 | 0.219 |
| D2  | 0.456 | -0.030 | 0.272 | D12 | 0.383 | -0.124 | 0.381 |
| D3  | 0.312 | -0.120 | 0.392 | D13 | 0.355 | -0.067 | 0.272 |
| D4  | 0.529 | 0.031  | 0.208 | D14 | 0.334 | -0.099 | 0.324 |
| D5  | 0.575 | -0.011 | 0.258 | D15 | 0.238 | -0.182 | 0.405 |
| D6  | 0.265 | -0.132 | 0.353 | D16 | 0.369 | -0.068 | 0.267 |
| D7  | 0.382 | -0.064 | 0.300 | D17 | 0.207 | -0.211 | 0.452 |
| D8  | 0.298 | -0.113 | 0.340 | D18 | 0.228 | -0.189 | 0.409 |
| D9  | 0.378 | -0.046 | 0.295 | D19 | 0.400 | -0.026 | 0.244 |
| D10 | 0.211 | -0.166 | 0.375 | D20 | 0.410 | -0.089 | 0.405 |